# A deep adversarial approach based on multi-sensor fusion for remaining useful life prognostics

DAVID VERSTRAETE[1], ENRIQUE DROGUETT[1,2], and MOHAMMAD MODARRES[1]

[1]*Department of Mechanical Engineering, University of Maryland, College Park, United States of America.*
*E-mail: dbverstr@terpmail.umd.edu*
*E-mail: modarres@umd.edu*
[2]*Department of Mechanical Engineering, University of Chile, Santiago, Chile.*
*E-mail: elopezdroguett@ing.uchile.cl*

Multi-sensor systems are proliferating the asset management industry and by proxy, the structural health management community. Asset managers are beginning to require a prognostics and health management system to predict and assess maintenance decisions. These systems handle big machinery data and multi-sensor fusion and integrate remaining useful life prognostic capabilities. We introduce a deep adversarial learning approach to damage prognostics. A non-Markovian variational inference-based model incorporating an adversarial training algorithm framework was developed. The proposed framework was applied to a public multi-sensor data set of turbofan engines to demonstrate its ability to predict remaining useful life. We find that using the deep adversarial based approach results in higher performing remaining useful life predictions.

## 1 Introduction

Reliability engineering has long been posed with the problem of predicting failures by using all data available. As modeling techniques have become more sophisticated, so too have the data sources from which reliability engineers can draw conclusions. The Internet of Things (IoT) and cheap sensing technologies have ushered in a new expansive set of multi-dimensional data which previous reliability engineering modeling techniques are unequipped to handle.

Diagnosis and prognosis of faults and remaining useful life (RUL) predictions with this new data are of great economic value as equipment customers are demanding the ability of the assets to diagnose faults and alert technicians when and where maintenance is needed (Si 2011). This new stream of data is often too costly and time consuming to justify labeling all of it. RUL predictions, being the most difficult, are also of the most value for the asset owner. They provide information for a state-of-the-art maintenance plan which reduces unscheduled maintenance costs by avoiding downtime and safety issues. Therefore, taking advantage of unsupervised learning-based methodologies would have the greatest economic benefit. Deep learning has emerged as a strong technique without the need for previous knowledge of relevant features on a labeled data set (Verstraete 2018). If faulty system states are unavailable or a small percentage of the fault data is labeled, deep generative modeling techniques have shown the ability to extract the underlying two-dimensional manifold capable of diagnosing faults.

Deep learning has been employed with success to remaining useful life estimation (RUL). Gugulothu (2017) employed a recurrent neural network (RNN) for RUL estimation. Malhotra (2016), Zhao (2016), Yuan (2016), Zheng (2017), Zhao (2017), Wu (2017), Aydin (2017), and Zhao (2017) all employ long short-term memory (LSTM) networks to estimate RUL. Ren (2017) incorporates feature extraction coupled with a deep neural network for RUL estimation. Li (2018) uses convolutional neural networks (CNN) and time-windowing to estimate RUL.

These previous works into RUL estimation do not attempt to develop an understanding of the underlying generative or inference model. Moreover, they used datasets which were fully labeled. Generative modeling provides the possibility to accomplish this without having to label what could be massive multi-dimensional noisy sensor data. Labeling this data would be costly and difficult. A valuable methodology would provide the flexibility to include a small percentage of labeled data as it becomes available.

To address these problems, this paper proposes the first algorithm which incorporates both variational and adversarial training for RUL prognostics. The novelty of this method has vast applications for fault diagnosis and prognosis. Furthermore, it can be incorporated for both new and existing system assets.

## 2 Background

### 2.1 *Generative Adversarial Networks*

Generative Adversarial networks (GANs) are a class of generative models where the density is learned implicitly via minimax game (Goodfellow 2014). This game's objective is to learn a generator distribution $P_G(x)$ identical to the real data distribution $P_{data}(x)$. When one does not necessarily want to explicitly obtain an inference model to diagnose a system fault and assign probability to every data $x$ in the distribution, GANs are a viable alternative. To accomplish this, the generator trains a neural network (NN) capable of generating the distribution $P_G(x)$ by transforming a vector of random noise variables, $P_{noise}(z)$. The generator's objective, $G(z)$, is trained by *playing* against an adversarial discriminator network parameterized by a separate neural network whose objective, $D(x)$, is to classify the data as real or fake. The optimal discriminator $D(x) = P_{data}(x)/[P_{data}(x) + P_G(x)]$ would ideally converge to the Nash Equilibrium (Nash 1950); however, there is no mechanism to control this. Formally, this value function is Eq. (1):

$$\min_G \max_D V(G, D) = \mathbb{E}_{x \sim P_{data}(x)}[\log(D(x)] + \mathbb{E}_{z \sim P_{noise}(z)}[\log(1 - D(G(z)))]. \quad (1)$$

where, $P_{data(x)}$ is the data distribution, $P_{noise(x)}$ is the noise distribution, $D(x)$ is the Discriminator objective function, and $G(z)$ is the generator objective function.

### 2.2 *Variational Autoencoders*

Variational autoencoders (VAEs) are a class of explicit generative models which yields both inference and generative models (Kingma &Welling, 2013). VAEs attempt to learn a model, $p(x|z)$, of latent variables, $z$, which generates the observed data, $x$. Commonly $p(x|z) \equiv p_\theta(x|z)$ is parameterized by a neural network with parameters $\theta$. For most cases the posterior distribution $p(z|x)$ is intractable. However, an approximate posterior distribution, $q_\phi(z|x)$, can be used to maximize the evidence lower bound (ELBO) on the marginal data log-likelihood. Formally, this is expressed as Eq. (2),

$$\log p(x) \geq \mathop{\mathbb{E}}_{q_\phi(z|x)}[\log p_\theta(x|z)] - \text{KL}(q_\phi(z|x)||p(z)) \quad (2)$$

From this, the objective is equivalent to minimizing the Kullbeck-Liebler (KL) divergence between $q_\phi(z|x)$ and $p(z|x)$. Note that $q_\phi(z|x)$ is usually parameterized by a neural network with parameters $\phi$. VAEs have been successfully applied to fault diagnosis problems in the recent past (San Martin 2018).

## 3 Proposed Framework

In this work, we propose a mathematical framework that encapsulates the following features: non-Markovian transitions for state space modeling (i.e., it is not assumed that all information regarding past observation is contained within the last system state), adversarial training mechanism on the training of the recognition $q_\phi(z_t|z_{1:t-1}, x_{1:t})$, variational Bayes for the inference and predictive model $p_\theta(x_t|x_{1:t-1}, z_{1:t})$, and adversarial variational filtering algorithm. We set $x_t$ as the observed sensor data, $z_t$ as the latent system state, $\phi_t$ is the recognition model parameters, $\theta_t$ is the inference model parameters, and $y_t$ is the target domain relevant to the adversarial training $y \in 0,1,\dots,RUL$.

We denote the latent sequence $z_t \in \mathcal{Z} \subset \mathbb{R}^{n_z}$ as a set of real numbers $n_z$. We denote observations $x_t \in \mathcal{X} \subset \mathbb{R}^{n_x}$ dependent on inputs $u_t \in U \subset \mathbb{R}^{n_u}$. Where $\mathcal{X}$ is potentially, but not limited to, a multi-dimensional data set consisting of multiple sensors from a physical asset. The observations themselves are not constrained to a Markovian transition assumption. Therefore, these transitions can be complex non-Markovian. This is often the case for engineering problems like crack growth and environmental effects on RUL. We are interested in the probabilistic function sequence $p(x_t|z_{1:t-1})$ generated by the discrete sequences $x_t = (x_1, x_2, \dots, x_t)$ and $z_{1:t-1} = (z_1, z_2, \dots, z_{t-1})$, as shown in Eq. (3).

$$p(x_t|x_{1:t-1}) = \int p(x_t|x_{1:t-1}, z_{1:t}) p(z_{1:t}|z_{1:t-1}) dz_{1:t} \quad (3)$$

$z_{1:t-1}, z_t \in \mathcal{Z} \subset \mathbb{R}^{n_z}$ denotes the latent sequence. The underlying latent dynamical system is assumed to have a generative model basis with emission model $p(x_t|x_{1:t-1}, z_{1:t})$ and transition model $p(z_t|z_{1:t-1})$. Two assumptions, Eq.'s (4) and (5) are classically imposed on emission and transition models to obtain the state space model,

$$p(x_t|x_{1:t-1}, z_{1:t}) = \prod_{i=1}^{t} p(x_t|z_t) \quad (4)$$

$$p(z_t|z_{1:t-1}) = \prod_{i=0}^{t-1} p(z_{t+1}|z_t) \quad (5)$$

It is assumed that the current state $z_t$ contains complete information for both the observations $x_t$, and the next state $z_{t+1}$. These assumptions are insufficient for complex non-Markovian transitions on noisy multi-dimensional sensor data. Therefore, we propose the objective function as shown in Eq. (6) which gives us an expressive approximate inference model $q_\phi(z_t|x_t)$. The mathematical formulation characterizes the state space model without assumptions as outlined in Eq.'s (2) and (3), and we also have both a generative and inference model of the system state to perform diagnostics and prognostics on the remaining useful life of the system.

$$\min_\theta \max_\phi \mathbb{E}_{D(x)} \mathbb{E}_{q_\phi(z_{1:t}|x_{1:t})} \big( [log p_\theta(x_{1:t}|z_{1:t})] - KL[q_\phi(z_{1:t}|x_{1:t}) \| p(z_{1:t})] \big) \tag{6}$$

This methodology is aided by GPU processing. Since this method does not include the Markov property, having to back propagate the biases and weights through each timestep is computationally expensive.

## 4 Experimental Results

To evaluate the proposed methodology the Commercial Modular Aero-Propulsion System Simulation (C-MAPPS) data set was used. CMAPPS is a tool developed and coded in MATLAB and Simulink environment for the simulation of commercial turbofan engines (Frederick 2007). The model takes an input parameter of an engine component degradation level or health indicator and outputs corresponding sensor signal values. Operational profile, closed-loop controllers and environmental conditions can all be adjusted to suit the specific problem the user is trying to solve. The 90,000-pound thrust class engine and the simulation package's flexibility allows operations at 1) altitudes ranging from sea level to 40,000 feet, 2) Mach numbers from 0 to 0.90, and 3) sea-level temperatures from -60 to 103 °F. The main elements of the engine are shown in Figure 1.

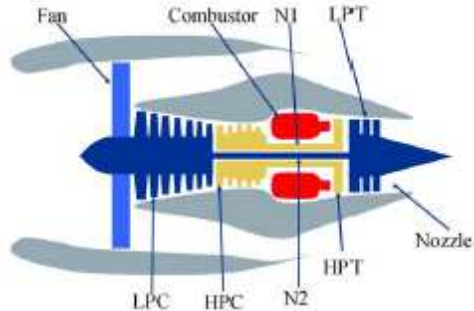

Figure 1: Simplified diagram of engine simulated in C-MAPPS (Frederick 2007).

Specifically, for this paper, FD001 of the PHM 2008 competition data set using CMAPPS is used for this analysis and application.

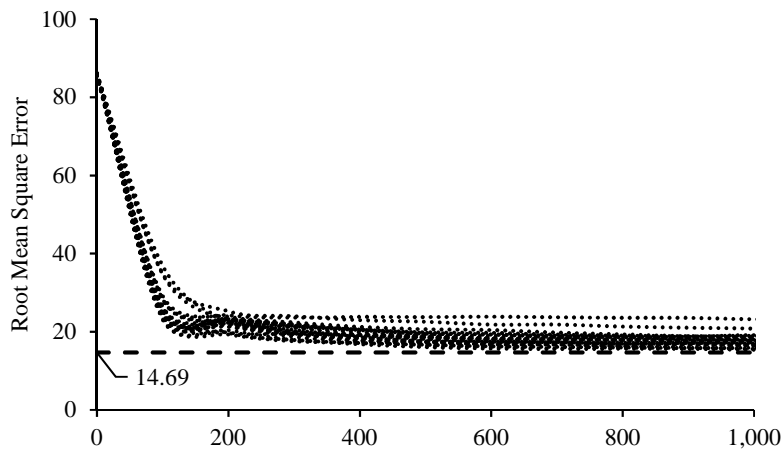

Figure 2: FD001 RMSE results vs training step for 50 iterations with the lowest result (14.69) marked.

The results from training fifty iterations and RUL estimations are a mean of 16.91 RMSE and a standard deviation of 1.48. The lowest result from the training was an RMSE of 14.69 as shown in Figure 2. These results are very good and near the state-of-the-art results for this data set. The output of the framework also includes a generative model that gives the engineer the ability to potentially generate more data. Moreover, these results are fully unsupervised learning, whereas similar results are fully supervised estimations (Li 2018). Further research will address these gaps and refine the results on a real-world application.

## 5  Conclusions

In this paper we have proposed a deep learning enabled adversarial-variational mathematical framework for remaining useful life estimation. Unsupervised RUL estimation is a critical area of structural health monitoring research. It has many applications into numerous industries. This mathematical formulation is the first application of its kind and shows great promise.

The proposed mathematical framework demonstrates a solid ability to predict the remaining useful life of the asset. An engineer can decide whether to plan for maintenance before a failure occurs and make the necessary arrangements. The application of the mathematical framework is not only limited to turbo-fan engines. Oil and gas, wind turbine farms, automotive, and aerospace can all benefit from this research.

### Acknowledgments

The authors acknowledge the partial financial support of the Chilean National Fund for Scientific and Technological Development (Fondecyt) under Grant No. 1160494.